\documentclass[
aps, prl, reprint,
superscriptaddress,
amsmath,amssymb,
floatfix,
]{revtex4-2}

\usepackage{color}

\newcommand\ddfrac[2]{\frac{\displaystyle #1}{\displaystyle #2}}

\usepackage{graphicx}
\usepackage{dcolumn}
\usepackage{bm}
\usepackage{enumitem}

\begin{document}

\title{Hidden multiscale organization and robustness of real multiplex networks}

\author{Gangmin Son}
\affiliation{Department of Physics, Korea Advanced Institute of Science and Technology, Daejeon 34141, Korea}

\author{Meesoon Ha}
\email[Corresponding author; ]{msha@chosun.ac.kr}
\affiliation{Department of Physics Education, Chosun University, Gwangju 61452, Korea}

\author{Hawoong Jeong}
\affiliation{Department of Physics, Korea Advanced Institute of Science and Technology, Daejeon 34141, Korea}
\affiliation{Center of Complex Systems, Korea Advanced Institute of Science and Technology, Daejeon 34141, Korea}

\date{\today}

\begin{abstract}

Hidden geometry enables the investigation of complex networks at different scales. Extending this framework to multiplex networks, we uncover a different kind of mesoscopic organization in real multiplex systems, named \textit{clan}, a group of nodes that preserve local geometric arrangements across layers. Furthermore, we reveal the intimate relationship between the unfolding of clan structure and mutual percolation against targeted attacks, leading to an ambivalent role of clans: making a system fragile yet less prone to complete shattering. Finally, we confirm the correlation between the multiscale nature of geometric organization and the overall robustness. Our findings expand the significance of hidden geometry in network function, while also highlighting potential pitfalls in evaluating and controlling catastrophic failure of multiplex systems.
\end{abstract}

\maketitle

\section{Introduction} 

Complex systems possess an intricate architecture that spans multiple scales. The network geometry paradigm paves the way for exploring the multiscale organization of complex networks~\cite{boguna_network_2021, garcia-perez_multiscale_2018,zheng_scaling_2021,villegas_laplacian_2023}. In particular, the concept of hidden metric spaces with hyperbolic geometry gives natural explanations for the common properties of real networks, such as degree heterogeneity, strong clustering, and small-world-ness~\cite{serrano_self-similarity_2008, krioukov_hyperbolic_2010,papadopoulos_popularity_2012}. Coarse graining of nodes based on their distances in a hidden metric space enriches the multiscale unfolding of networks~\cite{garcia-perez_multiscale_2018,zheng_scaling_2021}. For example, it allows studying self-similarity of the human connectome~\cite{zheng_geometric_2020}. However, the study of multiscale organizations has still been limited to single-layer networks.

Indeed, many real networked systems consist of multiple interdependent systems represented by multilayer or multiplex networks, which are of theoretical and practical significance due to intriguing phenomena not seen in single-layer networks~\cite{boccaletti_structure_2014,bianconi_multilayer_2018}. In multiplexes, if a node in one layer is attacked, its dependent nodes in the other layers break down as well. This interdependent nature can yield a catastrophic cascade of failures in mutual connectivity, which makes understanding the robustness of multiplex systems fascinating~\cite{buldyrev_catastrophic_2010,son_percolation_2012,baxter_avalanche_2012,gao_robustness_2011,dong_robustness_2013,bianconi_dangerous_2014,baxter_weak_2014,reis_avoiding_2014,baxter_correlated_2016,cellai_percolation_2013,min_network_2014,gross_dynamics_2023}.
In this context, recent publications have demonstrated the significance of hidden geometry~\cite{kleineberg_hidden_2016,kleineberg_geometric_2017}: in real multiplexes, geometric organization correlated across layers, which can enhance their robustness against targeted attacks. Nonetheless, previous studies have only focused on the lack of interlayer independence based on mutual information~\cite{kraskov_estimating_2004}. Therefore, essential questions remain: How are the layers correlated across a range of scales? How do the multiscale properties affect the robustness?

\begin{figure}[!b]
    \includegraphics[width=\columnwidth]{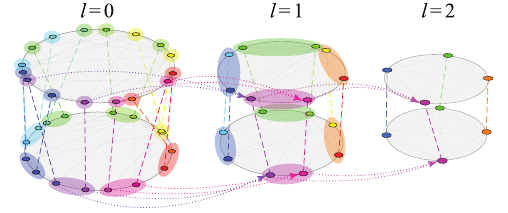}
    \caption{Multiscale unfolding of multiplex networks. The downscaled versions of a duplex are schematically illustrated as the zooming-out level $l$ increases, $l=0,1,2$ (from left to right). Each node has two angular coordinates for the upper and lower layers (gray disks). The colors of the nodes represent their angular coordinates in the lower layer, and dashed lines correspond to interlayer dependency links.
    }\label{fig1}
\end{figure}

In this paper, we show that the geometric correlations (GCs) of real multiplexes manifest across multiple scales rather than at a macroscopic scale. Notably, in contrast to the existing multiplex model for GCs~\cite{kleineberg_hidden_2016,kleineberg_geometric_2017}, real multiplexes exhibit the decrease of GCs as coarse graining. Our model with the mesoscopic groups of mutually close nodes, named \textit{clans}~\footnote{The term has been used in Ref.~\cite{ortiz_multiscale_2022} as a group of nodes with similar angular coordinates in the single-layer context.}, accounts for such nontrivial behaviors. Moreover, clan structure drastically affects the robustness against targeted attacks in an ambivalent way: the macroscopic organization between clans makes a system fragile, whereas the mesoscopic organization within clans constrains complete shattering at the end. These phenomena are elucidated based on the conceptual analogy between clan unfolding and mutual percolation in both real systems and our model. Finally, we confirm that the GC spectra predict the robustness stemming from intra-clan organization among diverse real multiplex systems.

\begin{figure*}[]
    \includegraphics[width=0.9\textwidth]{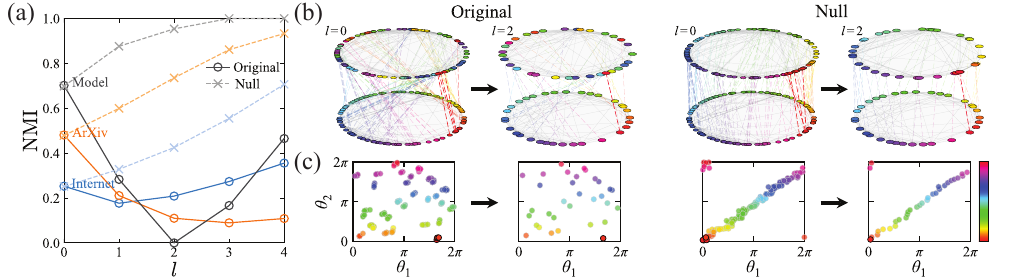}
    \caption{Geometric correlation (GC) spectra of real multiplexes and multiscale geometric multiplex model (MGMM). (a) The normalized mutual information (NMI)~\cite{kraskov_estimating_2004} as a function of zooming-out level $l$ (with the coarse-graining block size $\lambda=2$) for two sets of real data, i.e., arXiv (A48, orange circles) and Internet (I12, blue circles), and our model (MGMM, gray circles) with the total number of nodes, $N=2^{7}$, in the comparison with their null counterparts (crosses with lighter colors). (b) Multiscale unfolding of a synthetic multiplex generated by the MGMM and its null counterpart. The upper (lower) layer represents $\theta_1$ ($\theta_2$). In the original, nodes in a planted clan are highlighted (red dashed lines) for $l=0$ (left), which are coarse-grained into a single supernode for $l=2$ (right). In the null, the corresponding nodes are also highlighted. (c) $(\theta_1, \theta_2)$ space. Highlighted clans are also marked as bold black edges. The color of each node corresponds to $\theta_2$ in (b) and (c).}
     \label{fig2}
\end{figure*}

\section{Multiscale Unfolding\linebreak
of Multiplex Networks}

We start by extending the zooming-out technique of single-layer networks~\cite{garcia-perez_multiscale_2018} to multiplexes (see Fig.~\ref{fig1}). The approach relies on the assumption that each node in a network has radial and angular coordinates, $r_i$ and $\theta_i$, in a two-dimensional hyperbolic space~\cite{krioukov_hyperbolic_2010}. Since the radial coordinate $r_i$ reflects the expected degree of the node, $\kappa_i$, we only focus on angular coordinates $\{\theta_i\}$. Given a network with the angular coordinates of nodes and a block size $\lambda$, consecutive $\lambda$ nodes along the circle are grouped into a supernode whose angular coordinate $\phi$ is defined by
\begin{align}
    \xi e^{\phi} = \frac{1}{\lambda}\sum_{j=1}^{\lambda}{e^{i\theta_{j}}},
    \label{coarse-graining}
\end{align}
where $\theta_j$ is the angular coordinate of node $j$, and $\xi$ is the absolute value of the right hand side~\cite{faqeeh_characterizing_2018}. Extending this to multiplexes, the same mapping should be applied to every layer. Therefore, one chooses a standard layer to define a mapping. The iteration of this process yields a sequence of downscaled versions per multiplex (see Fig.~\ref{fig1}).

Measuring the GC~\cite{kleineberg_hidden_2016,kleineberg_geometric_2017} of downscaled versions yields a GC spectrum. For the sake of specificity, GCs were measured by the normalized mutual information (NMI)~\cite{kraskov_estimating_2004} between two sequences of angular coordinates in different layers; thus, we present the GC spectrum by the NMI as a function of the zooming-out level $l$ (see Supplemental Material (SM), Sec.~I~\cite{SM}). Here we investigate the GC spectra of real multiplexes (see SM, Sec.~II and Table~S1~\cite{SM}). Our aim is to compare real multiplexes with the existing model for GCs, called the geometric multiplex model (GMM). In the GMM, node $i$ at $\theta_{1,i}$ in layer 1 is assigned to $\theta_{2,i}=\theta_{1,i}+\Delta\theta_i$ in layer 2, where $\Delta\theta_i$ is an independent random variable. Thus, the GC is constructed at a macroscopic scale. To our aim, for a given multiplex, we obtain the GMM-like null counterpart, where the NMI for $l=0$ and the topologies of layers are the same, but dependency links are rearranged by independent local noise as in the GMM (see SM, Sec.~III~\cite{SM}).

Figure~\ref{fig2}(a) shows GC spectra for the arXiv collaboration (arXiv, A48) and the Internet (Internet, I12) multiplexes as well as the null counterparts with similar NMI values for $l=0$. Strikingly, we observe a significant discrepancy between the original and the null. In the null, GC spectra tend to increase monotonically, indicating that independent local noise is washed out as coarse graining. However, in the original, NMI values can decrease by zooming out. This kind of discrepancy is found in other real systems in our dataset (see SM, Table~S1~\cite{SM}), which can be quantified by the maximum difference as
\begin{align}
m=\max_{l}{\left[\text{NMI}_{\text{null}}(l) - \text{NMI}_{\text{org}}(l)\right]}.
\label{eq:m}
\end{align}

\section{Clan Structure}

To explain such nontrivial GC spectra in real multiplexes, we propose a multiplex model, named the multiscale geometric multiplex model (MGMM). Note that the NMI only indicates the lack of independence between two random variables, without specifying any particular correlation form, unlike the linear correlation coefficient, for instance. Therefore, a locally correlated yet globally uncorrelated configuration can also result in a nonzero NMI value. We introduce the groups of nodes preserving their local arrangement across layers, named \textit{clans}, to our model, the MGMM. Specifically, each group of consecutive $\Lambda$ nodes in layer~1 is defined as a clan; a node $i$ is assigned to an angular coordinate in layer~2, $\theta_{2,i}=\theta_{1,i}+\Delta\theta_{\text{clan}}$, where $\Delta\theta_{\text{clan}}$ is the same for nodes in the same clan. Finally, the angular arrangement within a clan is preserved, but between clans is totally randomized (see SM, Sec.~III~\cite{SM}). 

Figure~\ref{fig2}(b) schematically illustrates the MGMM and its GMM-like null counterpart with their downscaled versions, and in Fig.~\ref{fig2}(c), the MGMM with $\Lambda=2^2$ exhibits no macroscopic correlations but four nodes in a clan are close to each other across layers. Such local correlations lead to a nonzero NMI value at $l=0$ in Fig.~\ref{fig2}(a) (model, original). When each clan becomes a supernode at the zooming-out level $l=2$, the totally random organization between clans makes the downscaled version have no GCs. However, the GMM-like counterpart constructs a trivial linear correlation at a macroscopic scale, which leads to a monotonic increase in its GC spectrum. Consequently, our model with clans accounts for the nontrivial behavior of the GC spectra, not present in the existing model. Then a question arises: Does clan structure appear in real multiplexes?

\begin{figure}[!t] 
    \includegraphics[width=\columnwidth]{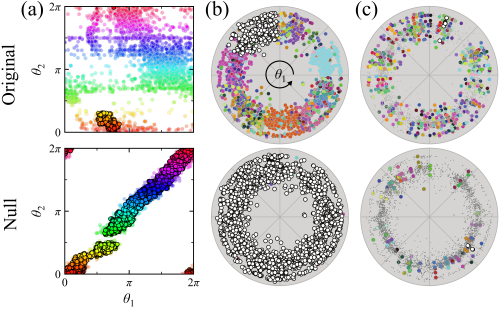}
    \caption{Multiscale organization of Internet. (a) Angular coordinates of nodes in ($\theta_1$, $\theta_2$) space for the Internet (top) and its null counterpart (bottom). The color of each node corresponds to $\theta_2$. Identified clans in the original Internet (top) and its null counterpart (bottom) for (b) $z=1/6$ ($\theta_w/\theta_c=5$) and (c) $z=2/3$ ($\theta_w/\theta_c=0.5$). The presented maps are for layer 1. Clan memberships correspond to colors, and if the clan size is less than $3$, nodes belonging to the clan are denoted as tiny gray dots. In particular, nodes in the largest clan are colored white and highlighted by bold black edges. Those in (b) are also highlighted in (a) in the same way.}
    \label{fig3}
\end{figure}

To answer the question, here we identify clans for a given multiplex. If the angular distance $d_{ij}$ between two nodes $i$ and $j$ is less than a certain angular window $\theta_{w}$, in both layers, they have the same clan membership.
Concretely, a characteristic scale $\theta_{c} = 2\pi \ln{N} / N$ among $N$ points randomly distributed on a unit circle~\cite{zuev_emergence_2015} allows us to define a resolution factor $z$ as
\begin{align}
    z=\ddfrac{1}{{1+{\theta_{w}}/{\theta_{c}}}}.
\end{align}
For $\theta_{w}=\infty$, $z=0$ and all the nodes belong to a single clan, and for $\theta_{w}=0$, $z=1$ and all the clans correspond to isolated nodes.
Figure~\ref{fig3} shows the identified clan structure of the Internet and its null counterpart. Although two multiplexes have the same GC at $l=0$ [see Fig.~\ref{fig2}(a)], the joint angular arrangements are clearly distinct from each other [Figs.~\ref{fig3}(a)]. As in the comparison of the MGMM with the GMM [see Fig.~\ref{fig2}(c)], in real multiplexes, layers seem uncorrelated at a macroscopic scale, while its null counterpart exhibits a clear linear correlation. This difference is reflected in the clan structure [Fig.~\ref{fig3}(b) and~\ref{fig3}(c)]. For $z=1/6$, in the original, plenty of mesoscopic clans appear, whereas, in the null, most nodes belong to a giant clan. For $z=2/3$, the null has more clans than the original, but most clans merely correspond to isolated nodes or pairs of nodes. Therefore, the nontrivial GC spectrum in Fig.~\ref{fig2}(a) results in the appearance of mesoscopic clans in real multiplexes.

\begin{figure}[]
    \includegraphics[width=\columnwidth]{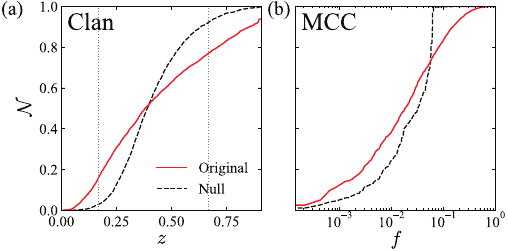}
    \caption{Clan unfolding and mutual percolation in Internet. The rescaled number of clusters $\mathcal{N}$ is plotted (a) for the clan against the resolution factor, $z$, and (b) for the mutually connected component (MCC) against the removal fraction of nodes, $f$, respectively. We compare the dynamics in the original multiplex (red solid lines) with its null counterpart (black dashed lines). The vertical gray dotted lines in (a) are drawn for $z=1/6$ and $z=2/3$ to indicate the instances in Fig.~\ref{fig3}.
    } 
    \label{fig4}
\end{figure}

The qualitative discrepancy of clan structure in Fig.~\ref{fig3} becomes apparent by the number of clans, $\mathcal{N}_{\text{clan}}$, as a function of $z$ in Fig.~\ref{fig4}(a). As expected in Fig.~\ref{fig3}, a reversal occurs between $z=1/6$ and $z=2/3$, indicating that clan structure in the original leads to an earlier appearance of mesoscopic clans that remain longer as $z$ increases. Such results for various real multiplexes support the presence of the mesoscopic clan structure in real multiplexes (see SM, Sec.~V and Figs.~S3--S6~\cite{SM}).

\section{Role of Clans in Robustness}

By definition, clans are simply connected components in an overlapped proximity network, which allows us to identify the analogy between clan unfolding and mutual percolation in multiplexes~\cite{buldyrev_catastrophic_2010,son_percolation_2012}.
First, the connection probability $p$ in the actual network is set as a function of the angular distance, $p \sim d^{1/T}$, where temperature $T$ controls the interaction range~\cite{krioukov_hyperbolic_2010}. Although the power-law form implies long-range connections, the limitation of $T \to 0$ makes the connection probability similar to that in the proximity network. Second, mutual percolation concerns mutually connected components (MCCs), defined by a similar but less stringent constraint compared to the components derived from overlapped edges. Third, the targeted attack strategy, i.e., the removal of the highest-degree nodes, especially resembles the removal of the longest edges, i.e., the increase of $z$ in clan unfolding. Specifically, the expected value of the average angular length of edges incident to a node with the expected degree $\kappa$ is given by
\begin{align}
    \int d(\theta, \theta') p(\theta, \kappa, \theta', \kappa') \text{d}\theta\text{d}\theta'\text{d}\kappa' \sim \log{\kappa}.
\end{align}
As a result, we conjecture that clan structure also plays an analogous role in mutual percolation against targeted attacks. Since our analysis controls macroscopic GCs, this notion alludes to the origins of the robustness of real multiplexes beyond Ref.~\cite{kleineberg_geometric_2017} (see SM, Table~S2~\cite{SM} for the summary of the analogy).

Figure~\ref{fig4}(b) shows the number of MCCs as a function of the removal fraction of nodes $f$ against targeted attacks. Remarkably, similarly to the results of clan unfolding in Fig.~\ref{fig4}(a), the relative order of $\mathcal{N}_{\text{MCC}}$ between the original and the null is reversed. However, the analogy is not complete, so the apparent reversal in mutual percolation is not common in real multiplexes. However, they tend to have the smaller $\mathcal{N_{\text{MCC}}}$, implying that clan structure impedes complete breakdown against targeted attacks (see SM, Sec.~V and Figs.~S7--S10~\cite{SM}).

\begin{figure}[!t]
    \includegraphics[width=\columnwidth]{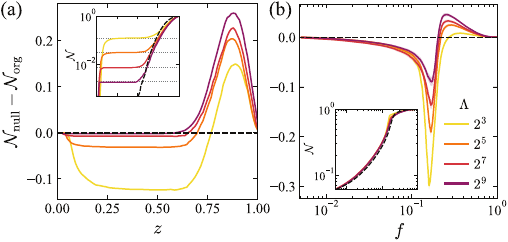}
    \caption{Clan unfolding and mutual percolation in MGMM. Synthetic multiplexes are generated by the MGMM~\cite{MGMM} for the total number of nodes, $N=2^{12}$, and the planted clan size $\Lambda \in\{2^3,~2^5,~2^7,~2^9\}$. The difference of $\mathcal{N}$ between the null (black dashed line) and the original instances (solid lines) are plotted (a) for the clan against $z$ and (b) for the MCC against $f$. Insets show the raw values of $\mathcal{N}$, and the horizontal gray dotted lines in the inset of (a) represent $1/\Lambda$ for each $\Lambda$.
    }
    \label{fig5}
\end{figure}

In order to systematically investigate the role of clans in mutual percolation, we employ synthetic networks generated by the MGMM for a variety of the planted clan size $\Lambda$. In Figs.~\ref{fig5}(a) and ~\ref{fig5}(b), we present $\mathcal{N}$ for clan unfolding and mutual percolation in synthetic networks, respectively. Given that GCs are similar to high NMI values ($\text{NMI}\approx 0.9$) as $\Lambda$ varies, we take a single null counterpart for them.
Notably, the crossing behaviors of the number of clans as $\Lambda$ varies [Fig.~\ref{fig5}(a)] are reflected in those of MCCs [Fig.~\ref{fig5}(b)], which demonstrates the ambivalent role of clans in percolation dynamics. In the MGMM, as $\Lambda$ increases, the size of planted clans grows and their number decreases, exposed as the plateaus in the inset of Fig.~\ref{fig5}(a), so the intra-clan organization becomes dominan over the inter-clan. Therefore, we find that for larger $\Lambda$, the crossing becomes less pronounced, but the final-stage robustness increases. Although the incompleteness of the analogy blurs the plateaus, the planted clan size $\Lambda$ plays a qualitatively similar role in both clan unfolding and mutual percolation (see SM, Sec.~V and Fig.~S11~\cite{SM}).

\begin{figure}[!t]
\includegraphics[width=\columnwidth]{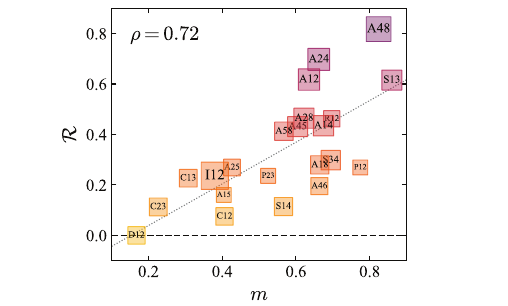}
    \caption{Correlation between $\mathcal{R}$ and $m$ for 22 real multiplexes (see SM, Table~S1~\cite{SM} for detailed information). The black dashed line indicates $\mathcal{R}=0$. The gray dotted line guides linear regression results. Square sizes correspond to the logarithm of system sizes, and colors to $\mathcal{R}$ for visual convenience.}
    \label{fig6}
    \end{figure}

Finally, from the implications of model results, we examine correlations between the nontrivial multiscale nature of geometric organization and robustness stemming from intra-clan organization in real systems. The multiscale nature of a multiplex can be quantified by the discrepancy in the GC spectrum with its null counterpart $m$ defined in Eq.~\eqref{eq:m}. The intra-clan robustness $\mathcal{R}$ can be defined by the suppression of complete shattering at the final stage observed in Figs.~4(b) and~5(b), as follows:
\begin{align}
\mathcal{R} = \max_{f}{\left[\mathcal{N_{\text{null}}} - \mathcal{N_{\text{org}}}\right]}.
\label{eq:R}
\end{align}
In other words, $\mathcal{R}$ describes how mesoscopic MCCs remaining after the removal of hubs are durable. In Fig.~\ref{fig6}, we find a strong positive correlation between the multiscale nature in GCs, $m$, and the robustness $\mathcal{R}$, (Pearson correlation coefficient $\rho \approx 0.72$ with the $\text{p-value} \approx 0.0002$). This supports our conjecture based on model results and emphasizes the significance of multiscale organization in percolation dynamics of real multiplexes.

\section{Conclusion}

To sum up, we filled the crucial gap between the existing multiplex model for geometric correlations (GCs)~\cite{kleineberg_hidden_2016,kleineberg_geometric_2017} and real multiplexes by hidden multiscale groups of mutually close nodes, i.e., \textit{clans}.
Remarkably, clans dictate the breakdown of mutual connectivity against targeted attacks, solely related to network topology, which highlights the power of the network geometry paradigm in elucidating network function through low-dimensional geometric patterns~\cite{van_emergence_2023}. This also implies that if clan structure is ignored in a multiplex, its robustness could be both over- and underestimated. Thus, the investigation of multiscale organizations has many applications to real systems~\cite{de_domenico_more_2023}, from the brain and power grids to physical materials~\cite{bonamassa_interdependent_2023}. The role of multiscale organizations on cascading failures~\cite{gross_dynamics_2023} is also a promising topic.

\begin{acknowledgments}
We would like to thank M.{\'A}. Serrano for helpful comments on the manuscript. This research was supported by the Basic Science Research Program through the National Research Foundation of Korea (NRF) (KR) [Grant No. NRF-2020R1A2C1007703~(G.S., M.H.) and No. NRF-2022R1A2B5B02001752~(G.S., H.J.)].
\end{acknowledgments}

\bibliography{main}

\begin{thebibliography}{40}
\bibitem [{MGM()}]{MGMM}%
  \BibitemOpen
  \href@noop {} {}\bibinfo {note} {Here the other parameters of the MGMM are set for a power-law degree distribution with the degree exponent $\gamma=2.6$, the average degree $\langle k \rangle \approx 8$, and temperature $T=0.7$.}\BibitemShut {Stop}%
\end{thebibliography}

\clearpage
\onecolumngrid

\begin{center}
\Large{\bf Supplemental Material}
\end{center}

\setcounter{equation}{0}
\setcounter{figure}{0}
\setcounter{table}{0}
\setcounter{section}{0}

\renewcommand{\theequation}{S\arabic{equation}}
\renewcommand{\thefigure}{S\arabic{figure}}
\renewcommand{\thetable}{S\arabic{table}}
\renewcommand{\thesection}{\Roman{section}}

\newcolumntype{L}[1]{>{\raggedright\arraybackslash}p{#1}} 
\newcolumntype{C}[1]{>{\centering\arraybackslash}p{#1}}
\newcolumntype{R}[1]{>{\raggedleft\arraybackslash}p{#1}}

\section{\label{sec:gcs} Geometric Correlation Spectrum}

Compared to the recent method for the multiscale unfolding of complex networks~\cite{garcia-perez_multiscale_2018}, we, here, introduce the simpler version, angular coarse-graining (ACG), inspired by the concept of angular coherence in Ref.~\cite{faqeeh_characterizing_2018}. Given a block size $\lambda$, the nearest $\lambda$ nodes are grouped into a supernode whose angular coordinate $\phi$ is defined by Eq.~(1) in the main text.

We apply the zooming-out method to multiplex networks. This method yields a sequence of downscaled replicas of a multiplex network as follows:
\begin{enumerate}
    \item Consider a duplex network with Layer~1 (L1) and Layer~2 (L2) with the given angular coordinates of its nodes at each layer.
    \item Obtain the angular coordinates of supernodes at L1 by applying the ACG with the block size $\lambda$, which gives a mapping between the original nodes and the supernodes at L1.
    \item Conduct the ACG based on the mapping defined in L1 to L2 so that the obtained supernodes are identical to those at L1.
    \item Iterate steps 2 and 3 for the obtained downscaled replica.
\end{enumerate}
Here we set $\lambda=2$, so the step of the ACG $l$ implies that $2^{l}$ nodes are mapped into a single supernode. In addition, the maximum step of the ACG could be reached to $\mathcal{O}(\log_{2}{N})$.

For each $l$ step, we can measure the angular correlation~\cite{kleineberg_hidden_2016,kleineberg_geometric_2017} between the layers in the downscaled replica. The angular correlation can be quantified by the normalized mutual information (NMI). Specifically, the NMI between two random variables $X$ and $Y$ can be written as
\begin{equation}
    \text{NMI}=\frac{I(X;Y)}{\max\{I(X;X), I(Y;Y)\}}
\end{equation}
where
\begin{equation}
    I(X;Y)=\int_{X}\int_{Y}{p(x,y)\ln{\left(\frac{p(x,y)}{p(x)p(y)}\right)}\text{d}x\text{d}y}
\end{equation}
is the mutual information between $X$ and $Y$ and $p(x,y)$ [or $p(x)$, $p(y)$] corresponds to the joint (or marginal) probability density function of $X$ and $Y$. 

To estimate the NMI value, we employ the method in Ref.~\cite{kraskov_estimating_2004} as the same as in Refs.~\cite{kleineberg_hidden_2016,kleineberg_geometric_2017}. Since a duplex gives two options for the standard layer, the average NMI value is used for geometric correlation spectra.

\section{\label{sec:dataset}Dataset}

We use a dataset for real-world multiplex networks~\cite{kleineberg_hidden_2016,kleineberg_geometric_2017,abdolhosseini-qomi_link_2020}. The largest mutually connected component (LMCC) of each multiplex network is considered to analyze its geometric organization and percolation dynamics. Since the network size should not be too small to apply multiscale unfolding, we only consider multiplex networks if the number of nodes in the LMCC is greater than or equal to $100$. The basic information of the selected cases is shown in Table~\ref{tab:real_mux}. The values of $m$ and $\mathcal{R}$ correspond to Fig.~6 in the main text.

\setlength{\tabcolsep}{12pt}
\begin{table}[!h]
    \centering
    \caption{Basic information of empirical multiplex networks}
    \begin{tabular}{lcrrrcc}
    \hline\hline
    Name & Abbreviation &LMCC & $|E_1|$& $|E_2|$& $m$& $\mathcal{R}$\\
    \hline
      Internet Layers 1, 2 &          I12 &  4710 & 24013 & 12683 & 0.380 & 0.236 \\
         ArXiv Layers 4, 8 &          A48 &  2252 &  7963 &  7285 & 0.824 & 0.817 \\
         ArXiv Layers 2, 4 &          A24 &   916 &  2607 &  3092 & 0.662 & 0.698 \\
         ArXiv Layers 1, 2 &          A12 &   790 &  2045 &  2141 & 0.635 & 0.618 \\
         ArXiv Layers 1, 4 &          A14 &   564 &  1540 &  1836 & 0.675 & 0.434 \\
         ArXiv Layers 2, 8 &          A28 &   521 &  1447 &  1479 & 0.622 & 0.464 \\
     SacchPomb Layers 1, 3 &          S13 &   510 &   805 &  1148 & 0.860 & 0.616 \\
         ArXiv Layers 4, 5 &          A45 &   506 &  1744 &  1388 & 0.605 & 0.431 \\
     SacchPomb Layers 3, 4 &          S34 &   426 &   839 &  1118 & 0.694 & 0.298 \\
         ArXiv Layers 5, 8 &          A58 &   310 &   826 &   907 & 0.567 & 0.413 \\
         ArXiv Layers 1, 8 &          A18 &   297 &   814 &   790 & 0.665 & 0.279 \\
     SacchPomb Layers 1, 4 &          S14 &   289 &   433 &   893 & 0.566 & 0.114 \\
    C. Elegans Layers 2, 3 &          C23 &   257 &   886 &  1561 & 0.226 & 0.113 \\
    C. Elegans Layers 1, 3 &          C13 &   247 &   512 &  1392 & 0.308 & 0.227 \\
    C. Elegans Layers 1, 2 &          C12 &   226 &   480 &   716 & 0.406 & 0.075 \\
    Drosophila Layers 1, 2 &          D12 &   222 &   347 &   324 & 0.167 & 0.000 \\
         ArXiv Layers 4, 6 &          A46 &   210 &   773 &   661 & 0.664 & 0.195 \\
         ArXiv Layers 2, 5 &          A25 &   182 &   477 &   429 & 0.426 & 0.269 \\
        Rattus Layers 1, 2 &          R12 &   158 &   234 &   183 & 0.697 & 0.462 \\
    Physicians Layers 2, 3 &          P23 &   106 &   230 &   181 & 0.525 & 0.236 \\
    Physicians Layers 1, 2 &          P12 &   104 &   226 &   226 & 0.775 & 0.269 \\
         ArXiv Layers 1, 5 &          A15 &   100 &   251 &   229 & 0.404 & 0.160 \\
    \hline\hline
    \end{tabular}
    \label{tab:real_mux}
\end{table}

\section{\label{sec:models}Models}

\subsection{Geometric Multiplex Model (GMM)}

Hidden hyperbolic geometry provides a natural explanation for the common properties of real-world networks, such as degree heterogeneity, clustering small-worldness, self-similarity, and navigability~\cite{serrano_self-similarity_2008,krioukov_hyperbolic_2010,boguna_navigability_2009}.
In this context, a simple model, called the $\mathbb{S}^1/\mathbb{H}^2$ model, has been proposed~\cite{serrano_self-similarity_2008,krioukov_hyperbolic_2010}.
In the formalism of the $\mathbb{S}^1$ model~\cite{serrano_self-similarity_2008}, each node $i$ has two hidden variables $\kappa_i$ corresponding to its expected degree and $\theta_i$ corresponding to its angular coordinate on a circle of radius $N/2\pi$, where $N$ is the total number of nodes.
Given $N$, the average degree $\bar{k}$, the degree exponent $\gamma>2$, and temperature $T\in[0,1)$, we generate a network instance for the $\mathbb{S}^1$ model as follows:

\begin{enumerate}

    \item Suppose that the probability density functions (PDFs) of $\theta$ are uniformly random and that of $\kappa$ is given by
    \begin{equation}
        \rho(\kappa) = (\gamma-1) \kappa_{\text{min}}^{\gamma-1} \kappa^{-\gamma},
    \end{equation}
    where $\kappa_{\text{min}}=\bar{k}(\gamma-2)/(\gamma-1)$ is the expected minimum node degree and sample the coordinates $\kappa_i$, $\theta_i$ of nodes $i = 1, \ldots, N$ from the PDFs.
    
    \item Connect each pair of nodes $i$, $j$ with probability
    \begin{equation}
    \label{connect_prob_S1}
    p(\kappa_i,\theta_i,\kappa_j,\theta_j)=\frac{1}{1+\left[\frac{d(\theta_i,\theta_j)}{\mu \kappa_i \kappa_j}\right]^{1/T}},
    \end{equation}
    where $d(\theta_i,\theta_j)=\frac{N}{2\pi}d_{ij}$ is the angular distance between nodes $i$,$j$ on the circle, $d_{ij}=|\pi-|\pi-|\theta_i-\theta_j|||$, and $\mu=\sin{T\pi}/2\bar{k}T\pi$.
\end{enumerate}

The equivalence between the $\mathbb{S}^1$ model and the $\mathbb{H}^2$ model can be shown by the relation between $\kappa_i$ and $r_i$
\begin{equation}
    r_i=R-2\ln{\frac{\kappa}{\kappa_{\text{min}}}},
\end{equation}
where $R$ is the radius of the hyperbolic disc in the $\mathbb{H}^2$ model with
\begin{equation}
        R=2\ln{\frac{N}{c}}\ \mbox{and}\
        c=\bar{k} \frac{\sin{T\pi}}{2T} \left(\frac{\gamma-2}{\gamma-1}\right)^2.
\end{equation}
We substitute the above relation into Eq.~\eqref{connect_prob_S1}, which leads to the connection probability in the $\mathbb{H}^2$ model
\begin{equation}
    p_{ij}=\frac{1}{1+e^{(x_{ij}-R)/2T}}
\label{eq:connection_prob}
\end{equation}
where $x_{ij} \approx r_i + r_j + 2\ln{(d_{ij}/2)}$ is the hyperbolic distance between nodes $i$ and $j$.

Conversely, both global parameters and hidden coordinates of the model can be inferred from a given network. The maximum likelihood estimation can be used to perform the inference problem~\cite{boguna_sustaining_2010,papadopoulos_network_2015,papadopoulos_network_2015-1}. Here we use the so-called mercator~\cite{garcia-perez_mercator_2019} to infer the hidden coordinates for a given network.

The network geometry paradigm has been extended to multiplexes~\cite{kleineberg_hidden_2016,kleineberg_geometric_2017}.
In a multiplex network, each layer can be embedded independently so that the coordinates for each layer are obtained.
Remarkably, it has been revealed that in real multiplex networks, the inferred coordinates for a layer are correlated with those of another layer.
In other words, real multiplex networks involve geometric correlations (GCs).
The radial correlation can be measured by Pearson correlation. The NMI can measure the angular correlation.

To generate synthetic networks with GCs, the GMM has been proposed~\cite{kleineberg_hidden_2016}.
In the GMM, each node $i$ is affiliated with two layers and has four hidden variables $\kappa_i^{(1)}$, $\theta_i^{(1)}$, $\kappa_i^{(2)}$, and $\theta_i^{(2)}$.
For the description of the GC, the hidden variables are generated with correlations.
The correlation between $\{\kappa_i^{(1)}\}$ and $\{\kappa_i^{(2)}\}$ is called radial correlation, reminiscent of degree correlation.
The correlation between $\{\theta_i^{(1)}\}$ and $\{\theta_i^{(2)}\}$ is called angular correlation, which can be interpreted as a kind of generalized version of community membership correlation.

The GMM constructs synthetic multiplex networks with GCs.
Specifically, each layer is constructed by using the $\mathbb{S}^1/\mathbb{H}^2$ model and the radial and angular coordinates are correlated across layers.
Here we only consider the GMM with two layers, where each node has four hidden variables: $\kappa_{i}^{(1)}$, $\theta_{i}^{(1)}$ at Layer~1 (L1) and $\kappa_{i}^{(2)}$, $\theta_{i}^{(2)}$ at Layer~2 (L2).

The detailed steps are as follows:
\begin{enumerate}
    \item Generate a network as L1 by the $\mathbb{S}^1$ model.
    \item Shuffle the angular coordinates of nodes based on truncated normal distribution,
    \begin{equation}
        f_{\text{trunc}}(d; g) \propto e^{-\frac{1}{2}(\tilde{g}d)^2},
    \end{equation}
    where $d$ is the angular distance between the original coordinate and the newly assigned coordinate, $\tilde{g}=g/(1-g)$ tunes the standard deviation with $g\in[0,1]$, and the condition of $d\in[0,\pi]$ limits the domain.
    
    \item Generate a network as L2 by the $\mathbb{S}^1$ model with the shuffled angular coordinates.
\end{enumerate}
Here we introduce a slight modification by using the circular normal (von Mises) distribution $f_{\text{circular}}(d; h) \propto e^{\tilde{h} \cos{d}}$ instead of the truncated normal distribution with $\tilde{h}=h/(1-h)$ and $h\in[0,1]$, which eliminates the need for an additional cutoff in the truncated normal distribution. In particular, $h=0$ corresponds to complete shuffling, and $h=1$ corresponds to the identical coordinates across layers.

\subsection{GMM-like Null Counterpart}

We propose a method to generate a null counterpart of a given multiplex network that yields the same NMI value between two angular coordinates but the angular displacements of interlayer dependency are independent as follows:
\begin{enumerate}
    \item Remove all the interlayer dependency links.
    \item Choose one node randomly in a layer and make it depend on a randomly chosen node in the other layer from $f_{\text{circular}}(d;h)$.
    \item Iterate step 2 until all nodes have their dependency links.
    \item Find the optimal value of $h$ which gives $\text{NMI}_{\text{null}} \approx \text{NMI}_{\text{org}}$ by numerically minimizing $(\text{NMI}_{\text{org}}-\text{NMI}_{\text{null}})^2$.
\end{enumerate}

\subsection{Multiscale Geometric Multiplex Model (MGMM)}

The MGMM constructs synthetic multiplex networks with V-shaped GC spectra.
In particular, when we assign the angular variables at Layer~2 (L2), the angular arrangement at Layer~1 (L1) is preserved for less than a specific scale $\Lambda$.
All the steps in the MGMM are the same as in the GMM except for the assignment of $\theta_{i}^{(2)}$.
So we can generate an instance in the MGMM as follows:

\begin{enumerate}
    \item Generate a network as L1 by the $\mathbb{S}^1$ model as in the GMM.
    \item Apply the ACG with the block size $\Lambda$ to L1.
    \item Shuffle the angular coordinates of supernodes from the circular normal distribution,
    \begin{equation}
        f_{\text{circular}}(d; h) \propto e^{\tilde{h} \cos{d}}.
    \end{equation}
    Here $d$ is the angular distance between the original coordinate and the newly assigned coordinate, $\tilde{h}=h/(1-h)$ tunes the standard deviation with $h\in[0,1]$.
    
    \item To generate L2, unwind the shuffled supernodes based on the relative angular coordinates of children nodes compared to the average, the angular coordinate of the supernodes.
    \item Generate a network as L2 by the $\mathbb{S}^1$ model with the shuffled angular coordinates.
\end{enumerate}

Unlike the shuffling of nodes in the GMM, for the shuffling of supernodes, two supernodes are randomly selected from $f_{\text{circular}}(d;h)$ with their distance $d$, and they are interchanged. This process is repeated for all supernodes at once. We adopt the interchange because it is the simplest method to maintain model consistency, ensuring a uniform distribution of angular coordinates. In the main text, we only consider $h=0$ for generating synthetic networks, so that the angular coordinates of supernodes (clans) are totally independent. In addition, $\Lambda=1$ makes the MGMM equivalent to the GMM.

\begin{figure}[]
    \includegraphics[width=\columnwidth]{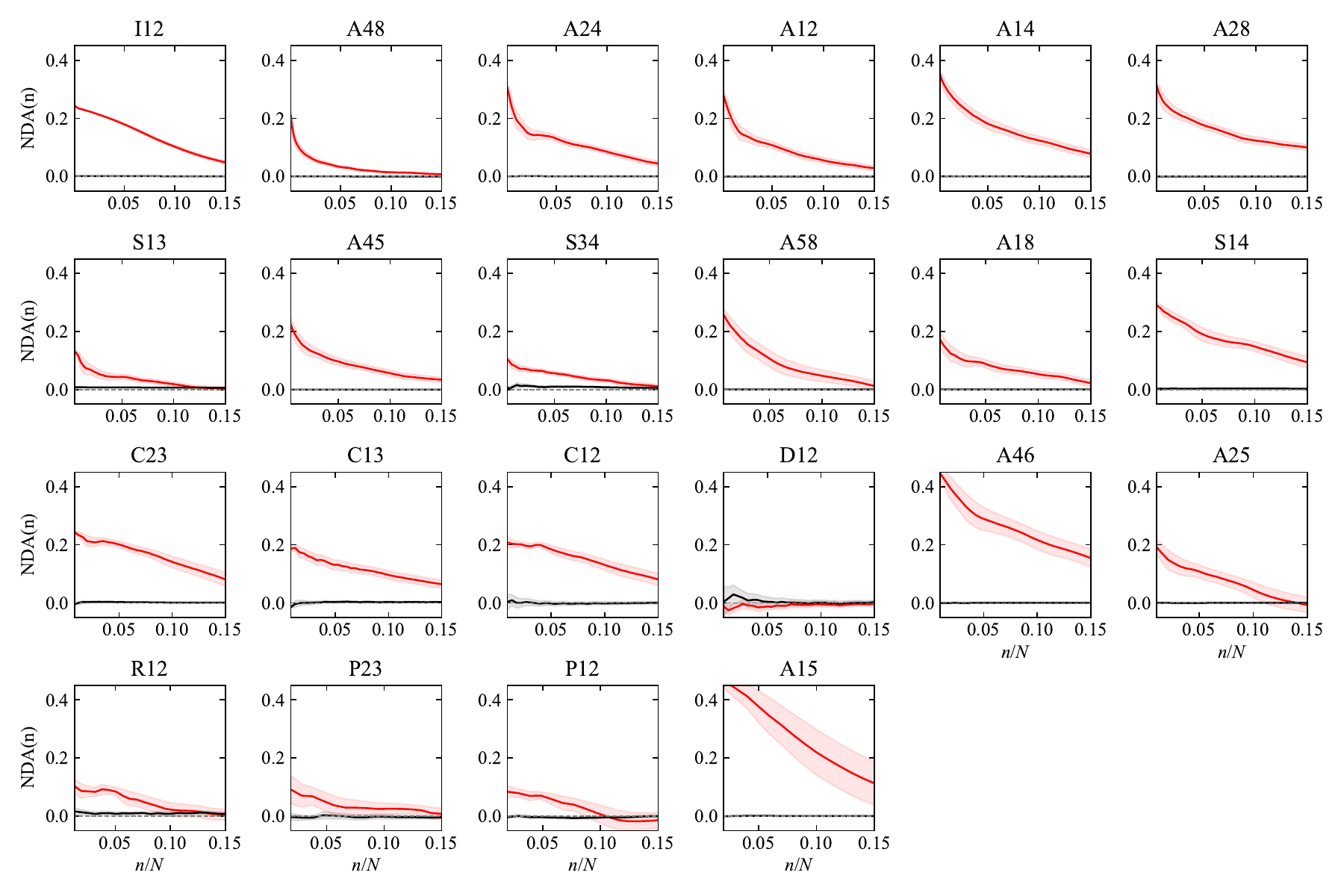}
    \caption{Normalized displacement alignment (NDA) as a function of $n/N$ for real multiplex networks (red) and their null counterparts (black).
    }\label{fig:NDA-empirical}
\end{figure}

\begin{figure}[]
    \includegraphics[width=0.65\columnwidth]{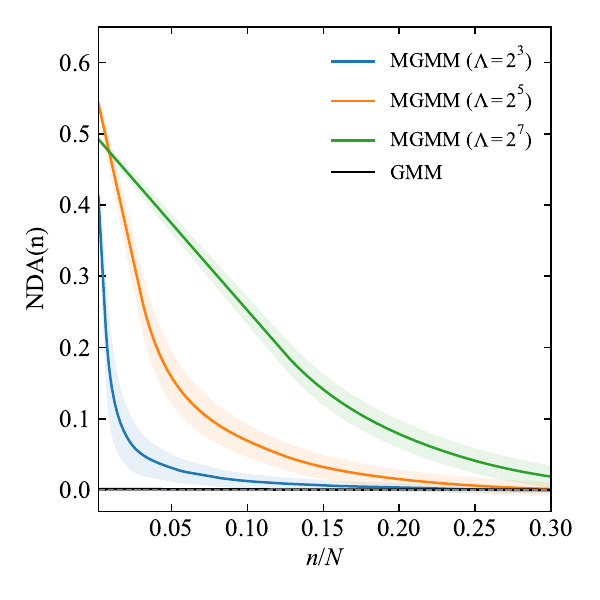}
    \caption{Normalized displacement alignment (NDA) as a function of $n/N$ for the MGMM (colored) and the GMM (black).
    }\label{fig:NDA-model}
\end{figure}

\section{\label{sec:dependency_displacement}Local Alignment of Dependency Displacement}

The main conclusion obtained by analyzing the geometric correlation spectrum is that in real multiplex networks, dependency displacements are locally aligned and the coarse-graining washes out the local alignment so that the geometric correlation decreases. In this section, we support this conclusion in terms of local alignment of dependency displacement.

We focus on the number of nodes that a dependency link passes through, rather than actual angular displacement. In order to do this, we assign $\phi_{i}^{(L)}$ to each node $i$ for each layer $L$ where $\phi_{i}^{(L)}=2\pi m/N$ and $m\in{1,\ldots,N}$, and $m$ is arbitrarily chosen with preserving the cyclic order of each layer. The dependency displacement of a node $i$ is defined as $\Delta\phi = \phi_{i}^{(1)} - \phi_{i}^{(2)} \in [-\pi,\pi]$. Then, the dependency alignment (DA) between a node and its $2n$ nearest nodes is defined as follows:
\begin{equation}
\text{DA}_i(n) = \frac{1}{2n}\sum_{\alpha\in\{-n,\ldots,n\}\setminus{\{0\}}}{1-d(\Delta \phi_{i}, \Delta \phi_{i+\alpha})/\pi}
\end{equation}
where $d$ is the angular distance, $n\in\{1,\ldots,\lfloor N/2\rfloor\}$, and the node index is ordered based on an arbitrarily chosen standard layer. Therefore, if node $i$ and its $2n$ nearest nodes completely preserve their relative positions, $\text{DA}_{i}(n)$ becomes $1$.
Let us denote the average value of DA as
\begin{equation}
    \text{DA}(n) = \frac{1}{N}\sum_{i=1}^{N}{\text{DA}_{i}(n)}.
\end{equation} We can define the normalized DA (NDA) as $\text{NDA}(n) = \text{DA}(n) - \text{DA}(\lfloor N/2\rfloor)$ to satisfy $\text{NDA}(\lfloor N/2\rfloor)=0$.

We measure the NDA for both empirical and synthetic multiplexes. Figure~\ref{fig:NDA-empirical} shows that in the empirical multiplexes, the values of the NDA are non-zero at $n/N\approx0$ and decrease; but their null counterparts exhibit $\text{NDA}(n)\approx0$ independent of $n/N$ (The only exception is D12, which exhibits a rather low value of $m$, indicating a minimal distinction from its GMM-like null counterpart). This implies that the dependency displacement of a node is correlated with its vicinity. This localized NDA is also shown in the comparison between the MGMM and the GMM (see Fig.~\ref{fig:NDA-model}).

\section{\label{sec:dataset}Role of Clan Structure in Mutual Percolation}

\begin{figure}[]
    \includegraphics[width=\columnwidth]{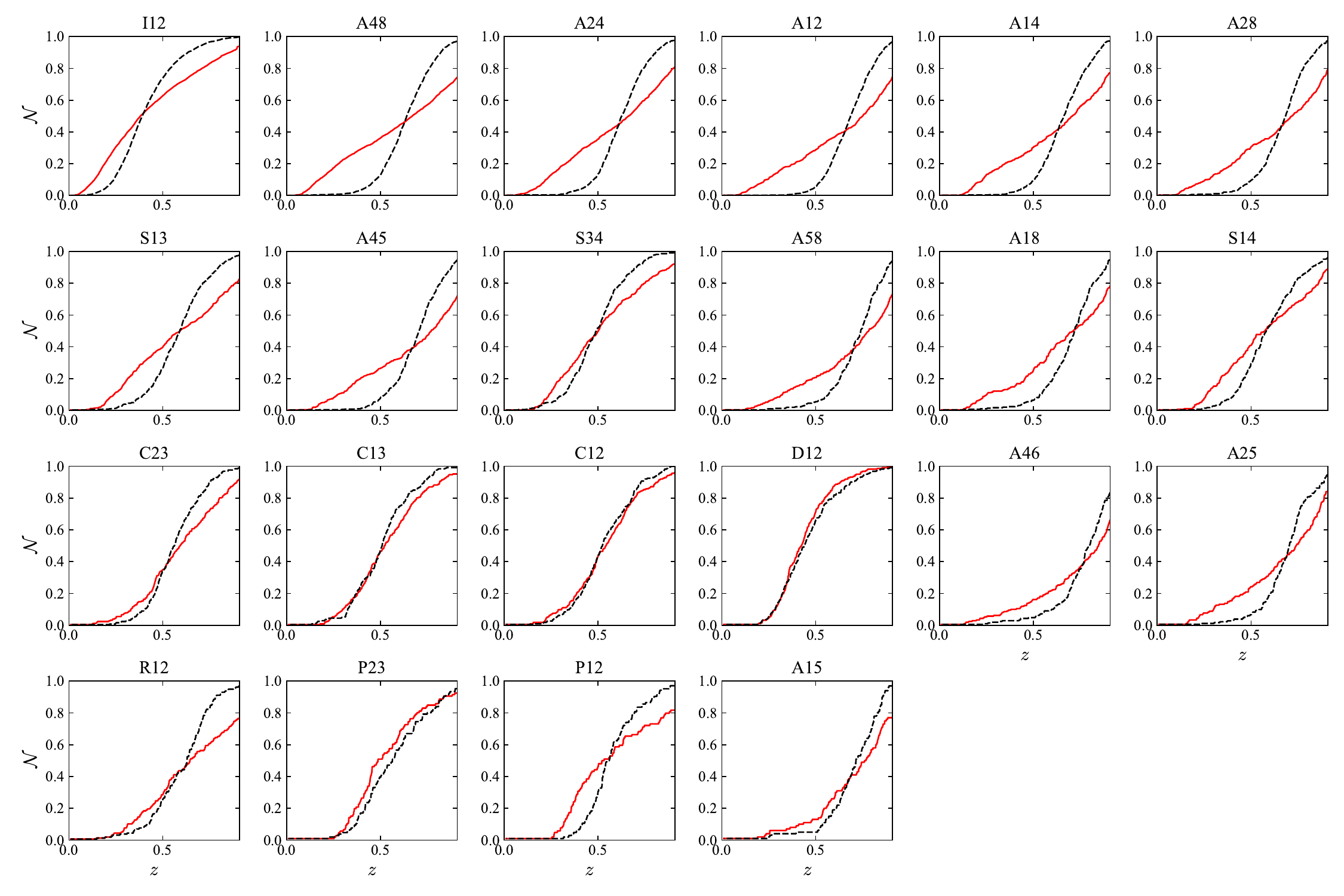}
    \caption{Number of clans as a function of resolution factor $z$ for real multiplex networks (solid red lines) and their null counterparts (dashed black lines).
    }\label{fig:empirical-clan-N}
\end{figure}

\begin{figure}[]
    \includegraphics[width=\columnwidth]{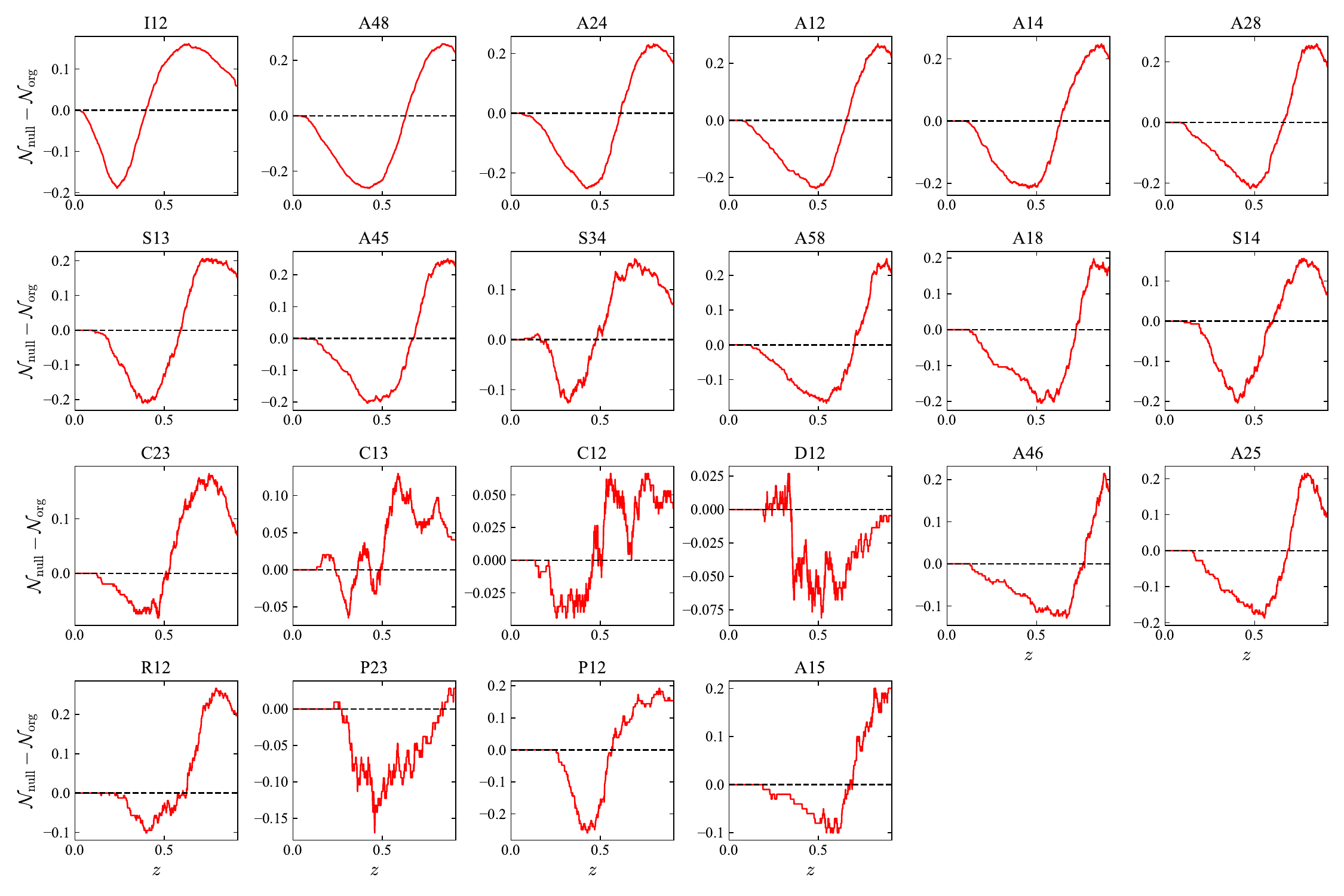}
    \caption{
    Difference of the number of clans as a function of resolution factor $z$ between real multiplex networks and their null counterparts.
    }\label{fig:empirical-clan-N_diff}
\end{figure}

As shown in Fig.~4 in the main text, the conceptual analogy between clan unfolding and mutual percolation, summarized in Table~\ref{tab:analogy}, leads to similar patterns in the number of clans and MCCs in the sense of comparing real multiplexes and their null counterparts as shown in Fig.~\ref{fig:empirical-clan-N}, \ref{fig:empirical-clan-N_diff}, \ref{fig:empirical-MCC-N}, and \ref{fig:empirical-MCC-N_diff}. 

\begin{figure}[]
    \includegraphics[width=\columnwidth]{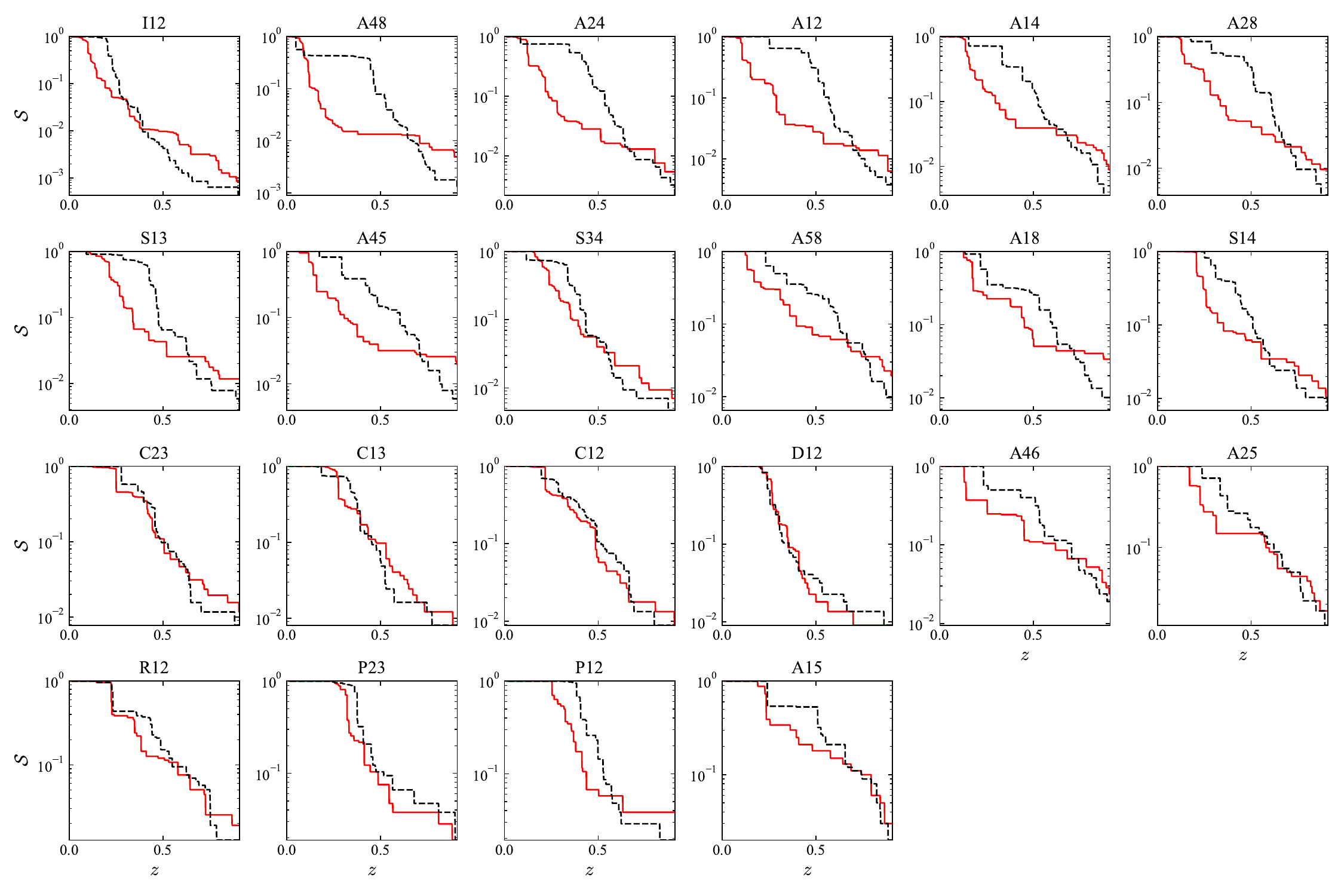}
    \caption{
    Largest clan size as a function of resolution factor $z$ for real multiplex networks (solid red lines) and their null counterparts (dashed black lines).
    }\label{fig:empirical-clan-S}
\end{figure}

\begin{figure}[]
    \includegraphics[width=\columnwidth]{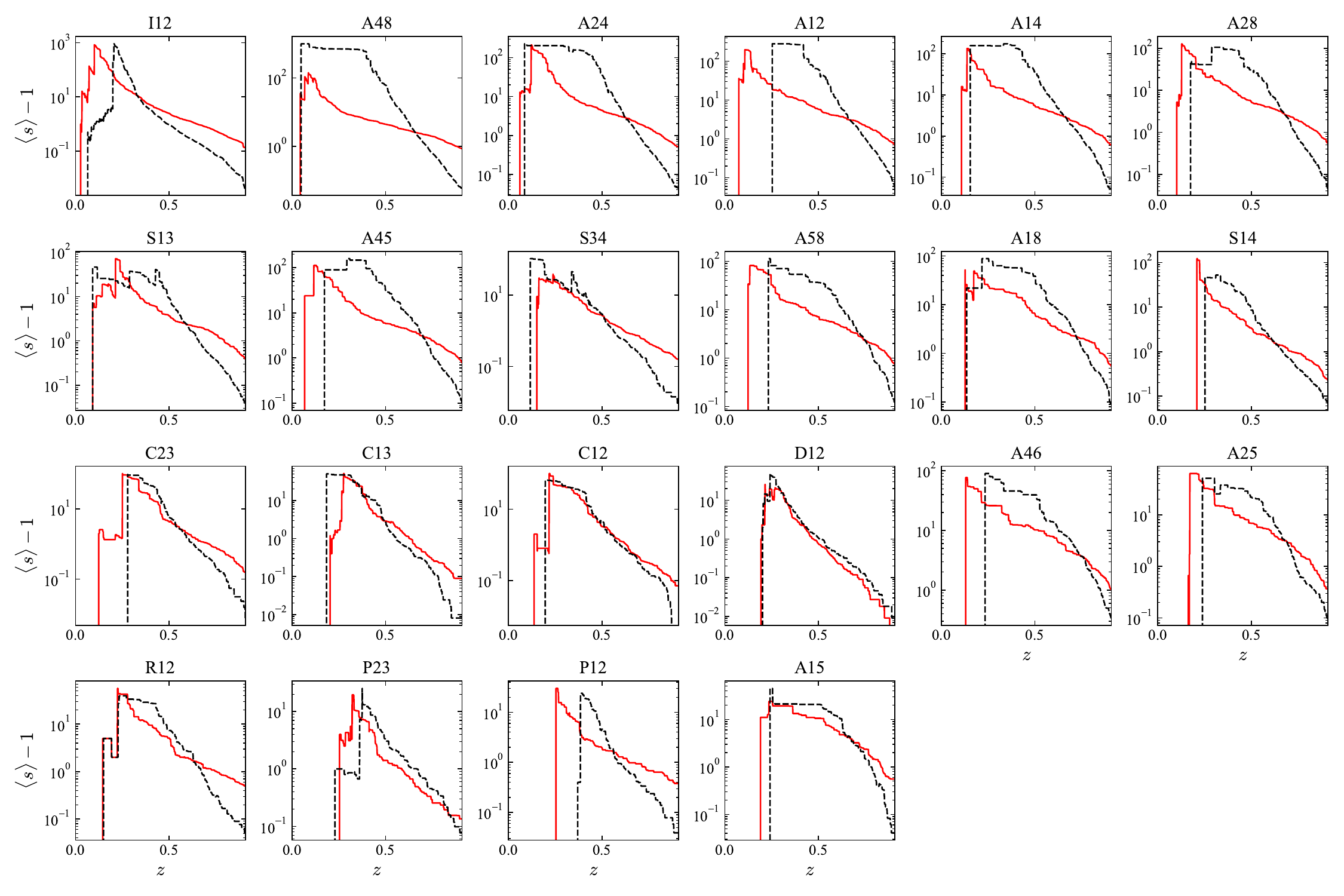}
    \caption{
    Average clan size as a function of resolution factor $z$ for real multiplex networks (solid red lines) and their null counterparts (dashed black lines).
    }\label{fig:empirical-clan-sus}
\end{figure}

\begin{figure}[]
    \includegraphics[width=\columnwidth]{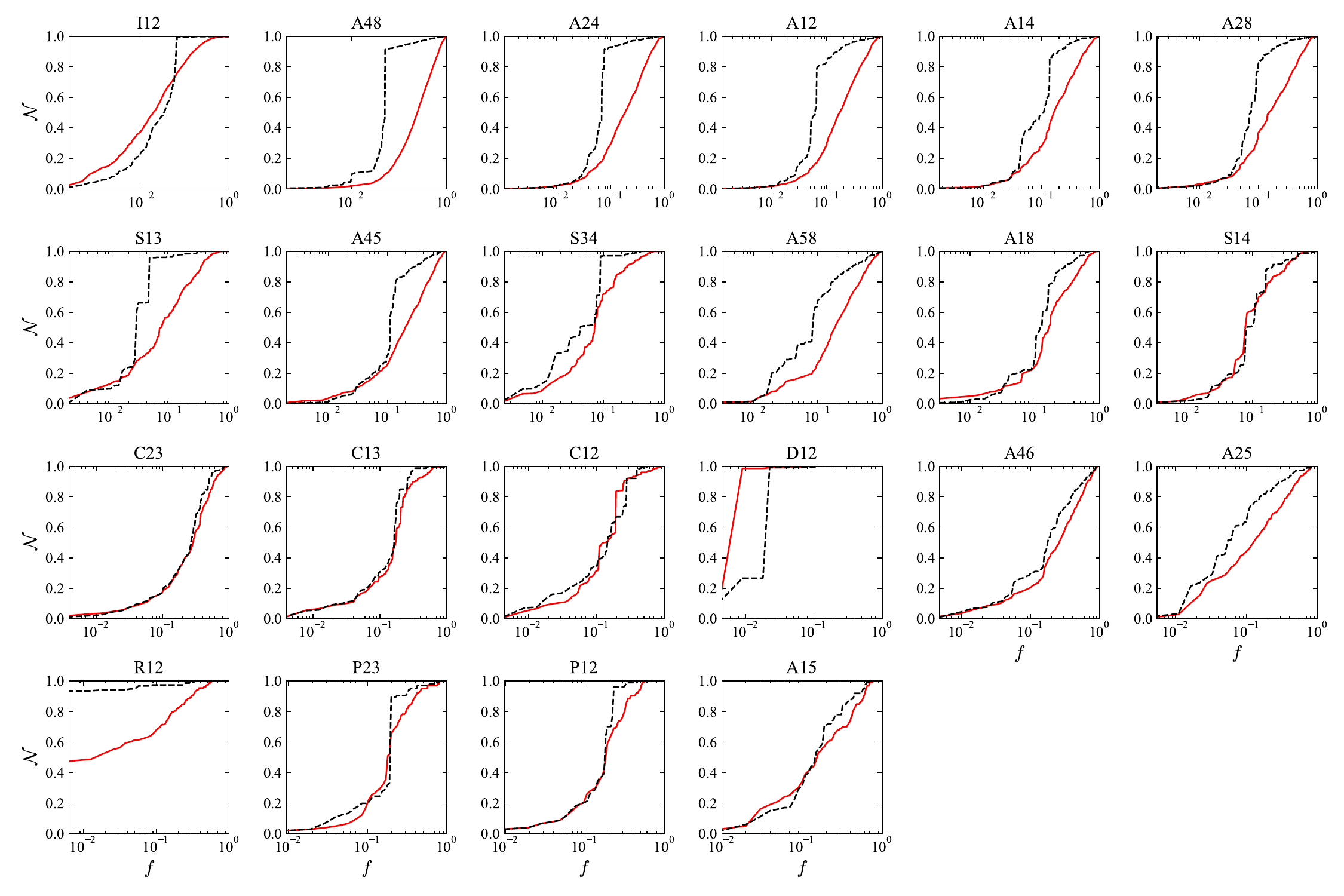}
    \caption{Number of MCCs as a function of removal fraction of nodes $f$ for real multiplex networks (solid red lines) and their null counterparts (dashed black lines).
    }\label{fig:empirical-MCC-N}
\end{figure}

\begin{figure}[]
    \includegraphics[width=\columnwidth]{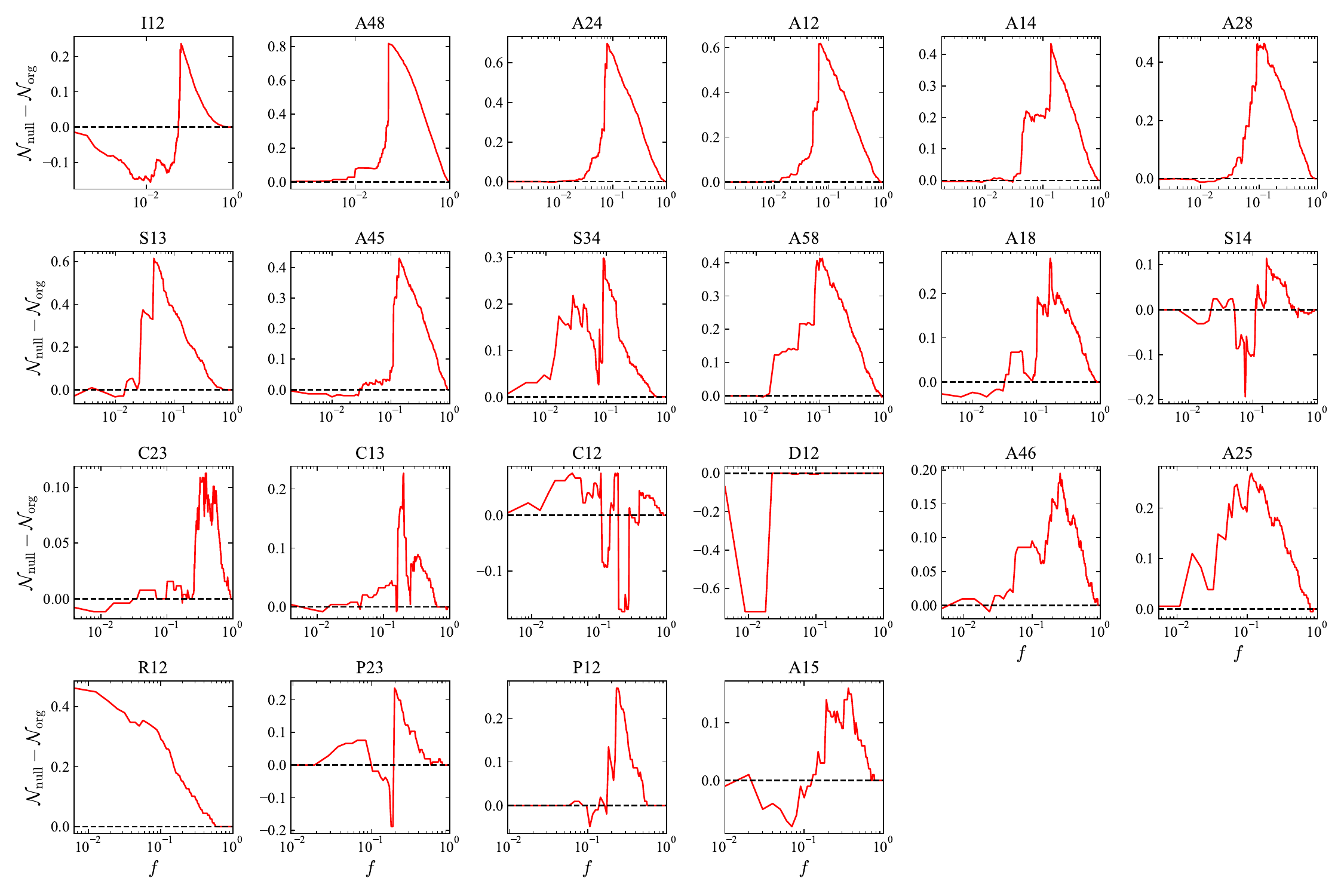}
    \caption{
    Difference of the number of MCCs as a function of removal fraction of nodes $f$ between real multiplex networks and their null counterparts.
    }\label{fig:empirical-MCC-N_diff}
\end{figure}

\begin{figure}[]
    \includegraphics[width=\columnwidth]{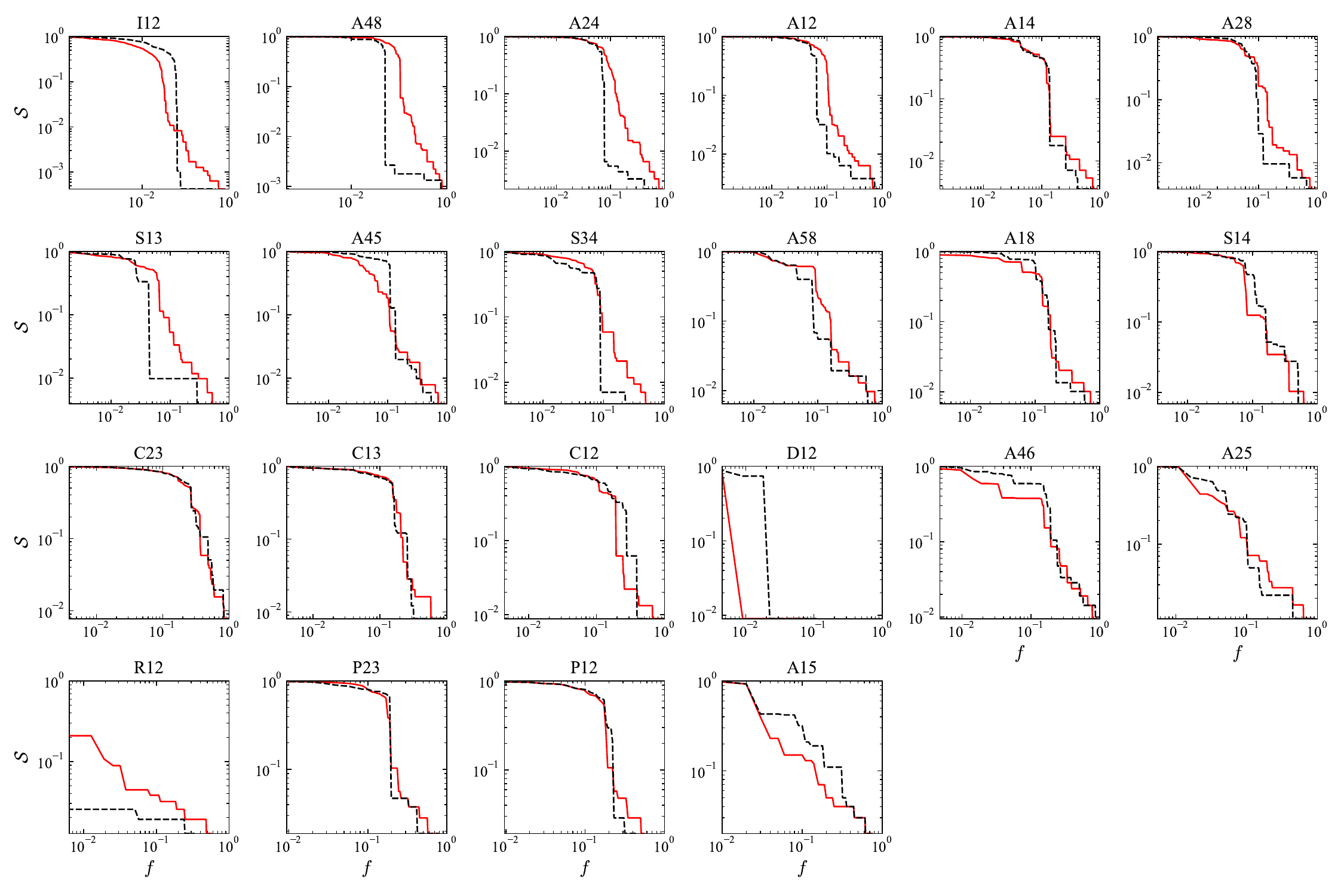}
    \caption{
    Largest MCC size as a function of removal fraction of nodes $f$ for real multiplex networks (solid red lines) and their null counterparts (dashed black lines).
    }\label{fig:empirical-MCC-S}
\end{figure}

\begin{figure}[]
    \includegraphics[width=\columnwidth]{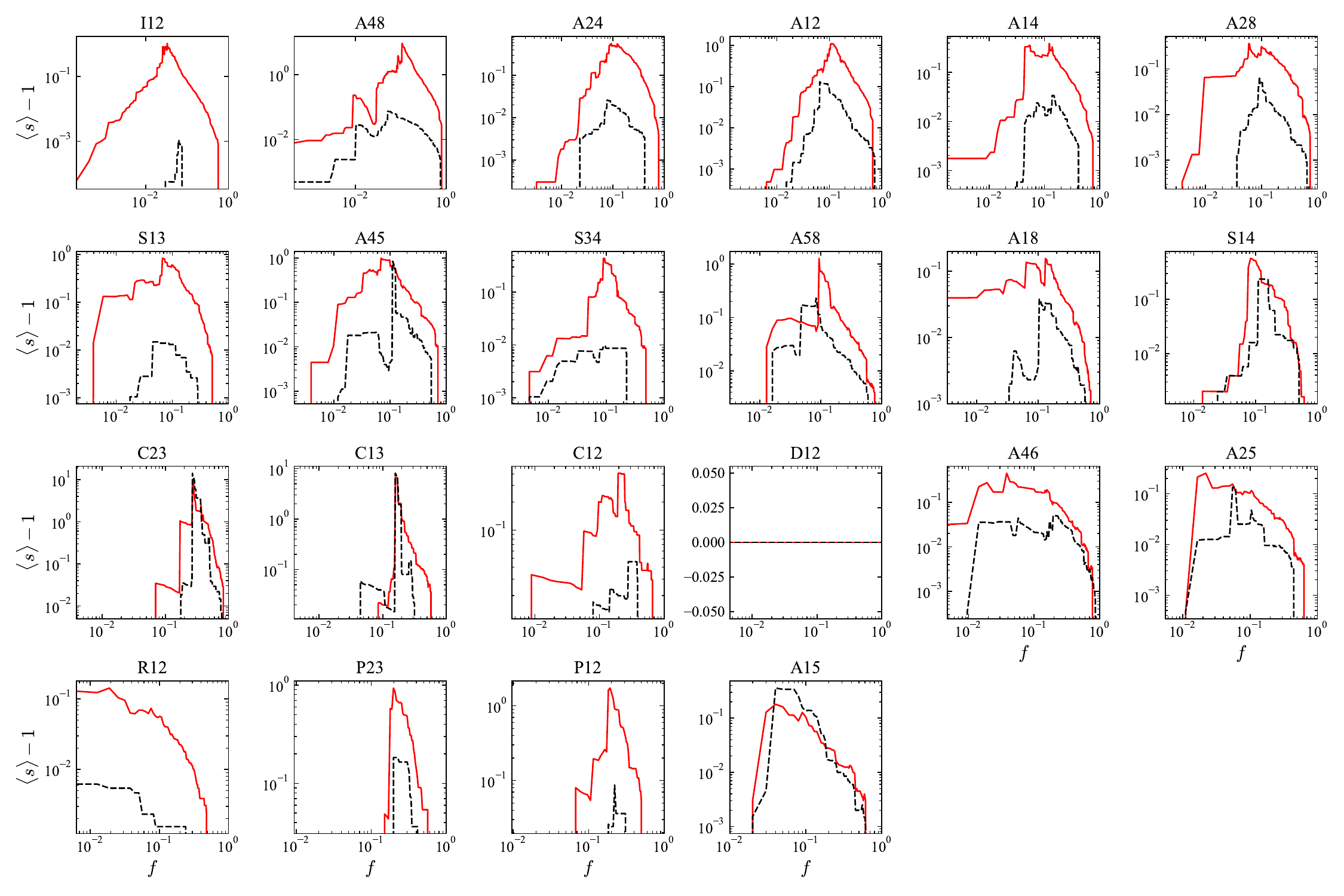}
    \caption{
    Average MCC size as a function of removal fraction of nodes $f$ for real multiplex networks (solid red lines) and their null counterparts (dashed black lines).
    }\label{fig:empirical-MCC-sus}
\end{figure}

Moreover, this finding is also supported by other quantities: the relative size of the largest cluster $\mathcal{S}$ and the average cluster size, often called the susceptibility, $\langle s \rangle = \sum{'}_{s}{s^{2} n_{s}} / \sum{'}_{s}{s n_{s}}$ where $n_{s}$ is the number of clusters with size $s$ and the primed sum excludes the largest cluster. 
First, a reversal occurs in the largest clan size $\mathcal{S_{\rm{clan}}}$ as shown in Fig.~\ref{fig:empirical-clan-S} in the opposite way of $\mathcal{N_{\rm{clan}}}$. However, similar to the number of MCCs $\mathcal{N_{\rm{MCC}}}$, the size of the largest MCC $\mathcal{S_{\rm{MCC}}}$ tends to show unclear crossing as shown in Fig.~~\ref{fig:empirical-MCC-S}. Second, as shown in Fig.~\ref{fig:empirical-clan-sus}, the average clan size shows earlier jumps for the original cases, indicating the earlier breakdown of the giant clan. In addition, the slower decay suggests that the mesoscale clans remain longer. These points also appear in the average MCC size as shown in Fig.~\ref{fig:empirical-MCC-sus}.

The prominent difference between clan and MCC in the average cluster size is that the original has a much higher peak than the null. This originates from the long-range connections in the actual networks. The absence of long-range connections in the proximity networks yields a ring along the angular axis. Therefore, the giant clan with $N$ nodes at $z=0$ breaks down by two initial angular gaps, thus leading to the trivial second-largest clan with the expected size $N/4$ and a peak at $\langle s \rangle \sim \mathcal{O}(N/4)$. Conversely, these trivial phenomena are absent in mutual percolation, so the role of clan structure is exposed as the higher peak of real multiplexes. 

\begin{figure}[]
    \includegraphics[width=0.75\columnwidth]{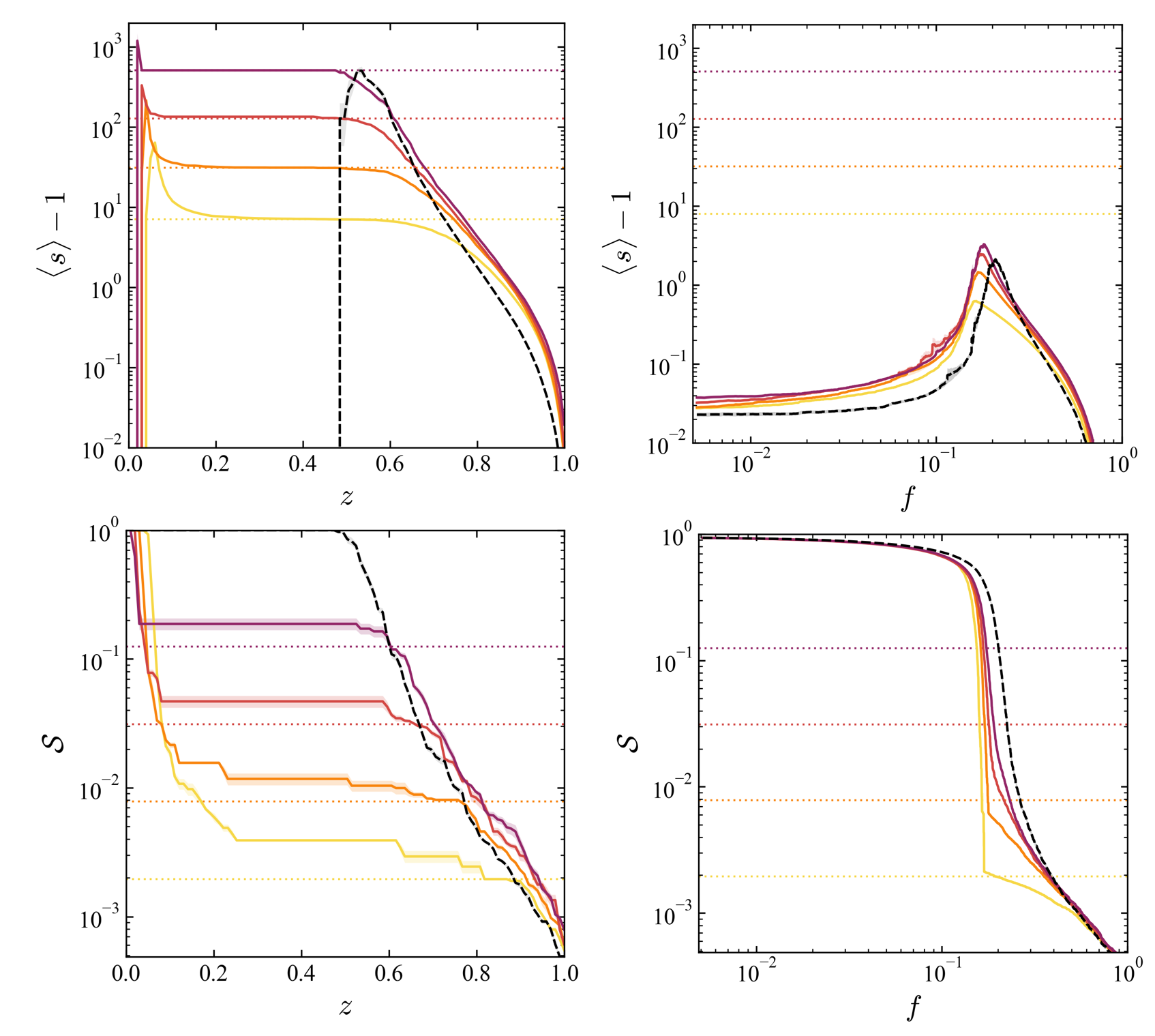}
    \caption{
    Average component size and largest component size of clan unfolding and mutual percolation for synthetic networks generated from MGMM (solid lines) and their null counterpart (dashed black lines). The dotted lines correspond to $\Lambda - 1$ in the top panels and $\Lambda / N$ in the bottom panels. The colors correspond to $\Lambda$ as the same as in Fig.~5 in the main text.
    }\label{fig:model}
\end{figure}

Finally, the comparison of the number of clans and MCCs for the MGMM in Fig.~5 in the main text is also supported by the largest cluster size and the average cluster size (see Fig.~\ref{fig:model}). For the clan unfolding, there appear plateaus dependent on the planted clan size $\Lambda$, which are blurred for the mutual percolation, as in Fig.~5. However, the increase of $\Lambda$ plays a qualitatively similar role in both cases.

\renewcommand{\arraystretch}{}
\begin{table}[!h]
    \centering
    \addtolength{\leftskip} {-2cm}
    \addtolength{\rightskip}{-2cm}
    \caption{Analogy between clan unfolding and mutual percolation}
    \begin{tabular}
    {L{1.5cm}C{7cm}C{7cm}}
    \hline\hline
     & Clan unfolding & Mutual percolation
     \\
    \hline
    
    Connection\newline
    probability &
    $p_{ij}=\Theta(\theta_w - d_{ij})$\footnote{$\Theta$ is the Heaviside step function.} &
    $p_{ij} \sim d_{ij}^{-1/T}$ (Eq.~\ref{eq:connection_prob})
    \\
    \hline
    
    Clusters &
    Connected components based on overlapped edges &
    Mutually connected components\newline
    (MCCs)
    \\
    \hline

    Removal strategy &
    Increase of the resolution factor $z$\newline
    ($=$ Removal of the longest edges) &
    Removal of the highest degree nodes\newline
    ($\approx$ Removal of the longest edges)
    \\
    \hline\hline
    \end{tabular}
    \label{tab:analogy}
\end{table}

\clearpage

\end{document}